\documentclass[10pt]{article}

 \usepackage{amsmath,amsthm,amssymb}
 \usepackage{makeidx,epsfig,lscape}
 \usepackage{color,colortbl}
 \usepackage{fancyhdr}
 \usepackage{xcolor,pict2e}

 \renewcommand{\headrulewidth}{0pt}
 \renewcommand{\footrulewidth}{0.5pt}

 \definecolor{myaqua}{rgb}{0.0,0.5,0.55}
 \definecolor{lightaqua}{rgb}{0.75,0.95,0.95}

 \usepackage[colorlinks = true,
            linkcolor = myaqua,
            urlcolor  = blue,
            citecolor = myaqua]{hyperref}

\usepackage{caption}
\usepackage{floatrow}
\def\lin#1#2{\textcolor[rgb]{0.6,0.6,0.6}{\vspace*{#1mm} \hrule
   height 3 pt \vspace*{#2mm}}}
\def\bt{\begin{tabular}}
\def\et{\end{tabular}}
\def\and{\mbox{ and }}

\def\1{{\bf 1}}

 \def\sectionn#1{\refstepcounter{section}{\color{myaqua}

 \vskip 6mm

 \noindent\Large\bf\thesection. #1}

 \vskip 3mm}

\begin{document}

 \fancyhead[L]{\hspace*{-13mm}
 \bt{l}{\bf Open Journal of *****, 2014, *,**}\\
 Published Online **** 2014 in SciRes.
 \href{http://www.scirp.org/journal/*****}{\color{blue}{\underline{\smash{http://www.scirp.org/journal/****}}}} \\
 \href{http://dx.doi.org/10.4236/****.2014.*****}{\color{blue}{\underline{\smash{http://dx.doi.org/10.4236/****.2014.*****}}}} \\
 \et}
 \fancyhead[R]{\includegraphics{pic1.ps}}

 $\mbox{ }$

 \vskip 12mm

{ 

{\noindent{\huge\bf\color{myaqua}
  Alignment of Quasar Polarizations on Large Scales Explained by Warped Cosmic Strings }}
%
\\[6mm]
{\large\bf Reinoud Jan Slagter$^1$}}
\\[2mm]
{ 
 $^1$ Asfyon, Astronomisch Fysisch Onderzoek Nederland
 and Department of Physics University of Amsterdam,
 The Netherlands\\
Email: \href{info@asfyon.com}{\color{blue}{\underline{\smash{info@asfyon.com}}}}\\[1mm]
 \\[4mm]
Received **** 2016
 \\[4mm]
Copyright \copyright \ 2014 by author(s) and Scientific Research Publishing Inc. \\
This work is licensed under the Creative Commons Attribution International License (CC BY). \\
\href{http://creativecommons.org/licenses/by/4.0/}{\color{blue}{\underline{\smash{http://creativecommons.org/licenses/by/4.0/}}}}\\
 \includegraphics{pic2.ps}

\lin{5}{7}
 { 
 {\noindent{\large\bf\color{myaqua} Abstract}{\bf \\[3mm]
 \textup{The recently discovered  alignment of quasar polarizations on very large scales could possibly explained by considering cosmic strings on a warped five dimensional spacetime.
Compact objects, such as cosmic strings, could have tremendous mass in the bulk, while their warped manifestations in the brane can be consistent with general relativity in 4D.
The self-gravitating cosmic string induces gravitational wavelike disturbances which could have effects felt on the brane, i.e., the massive effective 4D modes (Kaluza-Klein modes) of the perturbative 5D graviton.
This effect is amplified by the time dependent part of the warp factor. Due to this warp factor, disturbances don't fade away during the expansion of the universe.
From a non-linear perturbation analysis it is found that the effective Einstein 4D equations on an axially symmetric spacetime, contain a "back-reaction" term on the righthand side  caused by the projected 5D Weyl tensor and can act as a dark energy term.
The propagation equations to first order for the metric components and scalar-gauge fields  contain  $\varphi$-dependent terms, so the approximate wave solutions are no longer axially symmetric. The disturbances, amplified by the warp factor, can possess  extremal values for fixed polar angles. This could explain the two preferred polarization vectors mod $(\varphi, 90^o)$.}}}
 \\[4mm]
 {\noindent{\large\bf\color{myaqua} Keywords}{\bf \\[3mm]
quasar polarization -- cosmic strings -- warped brane world models -- U(1) scalar-gauge field -- multiple-scale analysis}
 \fancyfoot[L]{{\noindent{\color{myaqua}{\bf How to cite this
 paper:}} R. J. Slagter (2016)
  Alignment of Quasar Polarizations on Large Scales Explained by Warped Cosmic Strings}}
\lin{3}{1}
\renewcommand{\headrulewidth}{0.5pt}
\renewcommand{\footrulewidth}{0pt}
 \pagestyle{fancy}
 \fancyfoot{}
 \fancyhead{} 
 \fancyhf{}
 \fancyhead[RO]{\leavevmode \put(-90,0){\color{myaqua}R. J. Slagter}}
 \fancyhead[LE]{\leavevmode \put(0,0){\color{myaqua}R. J. Slagter} }
 \fancyfoot[C]{\leavevmode
 \put(-2.5,-3){\color{myaqua}\thepage}}
 \renewcommand{\headrule}{\hbox to\headwidth{\color{myaqua}\leaders\hrule height \headrulewidth\hfill}}
\sectionn{Introduction}\label{Intro}
{ \fontfamily{times}\selectfont
 \noindent Physicists speculate that extra spatial dimensions could exist in addition to our ordinary 4-dimensional spacetime. The underlying theory is the string theory, an unified description of gauge interactions and gravity.
String theory could provide an adequate description of quantum gravity and  can be used to explain the several shortcomings of the Standard Model and modern cosmology, i.e., the unknown origin of dark energy and dark matter, the weakness of gravity ( hierarchy problem) and the  incredibly fine-tuning of the cosmological constant. Moreover, the recently found evidence for the acceleration of our universe could be explained in these so-called super-string models without the need for a cosmological constant ( self-acceleration). However, its weak point is, that it is extremely hard to make predictions which are testable at energies available in experiments because the theory will manifest itself at energies of the order of the fundamental Planck scale $M_{Pl}$, dependent of the number of the extra dimensions.
The observed 4-dimensional Planck scale is given by Newton's constant and is $M_4\approx 10^{18} GeV$.
String theory also predicts the existence of sub-manifolds of the "bulk" spacetime, the so-called branes: it may be that our (3+1)-dimensional spacetime is such a 3-brane.  All standard model fields resides on the brane, while gravity can propagate into the bulk. The fundamental scale in elementary particle physics, the electroweak scale, is of order  $E_{EW}\approx 10^3 GeV$.
In order to lower down the fundamental scale of super string theory to the electroweak scale, one conjectures that the 4-dimensional Planck scale is not fundamental, but only an effective scale which can become much larger than the $M_{Pl}$ if the extra dimensions $L$ are much larger than $M_{Pl}^{-1}$\cite{arkani98}. For $L\sim 1$ mm, the fundamental Planck scale can be of the order of the electroweak scale.
Within this brane world picture, at low energies, gravity is localized at the brane and general relativity is recovered, but at high energy gravity "leaks" into the bulk. Recently there is growing interest in the warped brane world model\cite{randall99}, where there is one preferred extra dimension, with other extra dimensions treated as ignorable. The extra dimension is curved ( or warped) rather than flat. This means that self-gravity of the brane is incorporated.

It is conjectured that one needs an inflaton field in the very early stages of our universe to solve the problems in the standard model of cosmology, i.e., the horizon and flatness problem.  The inflationary cold dark matter model with a cosmological constant ($\Lambda$CDM) could be a good candidate if one abandons the cosmological constant problems. It can  explain the fluctuations we observe in the cosmic microwave background (CMB). The inflaton field could be the well-known scalar-Higgs field. This field has lived up to its reputation.
It originates from the theory of type II superconductivity, where vortex lines occur as topological defects in an abelian U(1) gauge model, which is coupled to a charged scalar field. It explains the famous Meissner effect (Ginzburg-Landau theory). Topological defects can occur when the field symmetries are broken. In cosmology, this happens when the universe cools down\cite{vilenkin94}.
Topological defects, such as cosmic strings, monopoles and textures, can have cosmological implications. Apart from their possible astrophysical roles, topological defects are fascinating objects in their own right and can give rise to a rich variety of unusual phenomena.
The  U(1) vortex solution possesses mass, so  it will couple to gravity. It came as a big surprise that there exists vortex-like solutions in general relativity. It is conjectured that any field theory which admits cosmic string solutions, a network of strings inevitable forms at some point during the early universe.However, it is doubtful if they will persist to the present time in the $\Lambda$CMD model.
Evidence of these objects would give us information at very high energies in the early stages of the universe.
It is believed that the grand unification (GUT) energy scale of symmetry breaking $\eta$ is about $10^{16}GeV$. The thickness of a cosmic string is $\sim \eta^{-1}\approx 10^{-30} cm$ and the length could be unbounded long.
The mass per unit length of a cosmic sting will be of the order of $10^{18}$kg per cm, which is proportional to the square of the energy breaking scale.  The thickness is still a point of discussion. By treating the cosmic string as an infinite thin mass distribution, one will encounter serious problems in general relativity. This infinite thin string model give rise to the "scaling solution", i.e., a scale-invariant spectrum of density fluctuations, which in turn leads to a scale invariant  distribution of galaxies and clusters. It was believed that  cosmic strings could have served as seeds for the formation of galaxies.
Cosmic strings can collide with each other and will intercommute to form loops. These loops will oscillate and loose energy via gravitational radiation and decay. There are already tight constraints  on the gravitational wave signatures  due to string loops via observations of the millisecond pulsar-timing data, the  cosmic  background radiation (CMB) by LISA and analysis of data of the LIGO-Virgo gravitational-wave detector. Its spectrum will depend on the string mass $G\mu$, where $\mu$ is the mass per unit length. Recent observations from the COBE, Wamp and Planck satellites put the value of $G\mu <10^{-7}$. It turns out that cosmic strings can not provide a satisfactory explanation for the magnitude of the initial density perturbations from which galaxies and clusters grew. The interest in cosmic strings faded away, mainly  because of the inconsistencies with the power spectrum of the CMB. Moreover, they will produce a very special pattern of lensing effect, not found yet by observations.
New interest in cosmic strings arises when it was realized that cosmic strings could be produced within the framework of string theory inspired cosmological models. Investigations on cosmic strings in warped brane world models show consistency with the observational bounds\cite{slagter14,slagter15}. The warp factor makes these strings consistent with the predicted mass per unit length on the brane, while brane fluctuations can be formed dynamically due to the modified energy-momentum tensor components of the scalar-gauge field. This effect is triggered by the time-dependent warp factor.
The recently discovered "spooky" alignment of quasar polarization over a very large scale\cite{hutsemekers14} good be well understood by the features of the cosmic strings in brane world models and could be the first evidence of the existence of these strings.

In section 2 we will outline the warped 5-dimensional model. In section 3 we apply  the multiple-scale approximation in order to find an wave-like solution to first order of the Einstein and matter field equations.
\sectionn{The warped 5D model}\label{5D model}
{ \fontfamily{times}\selectfont
 \noindent
Let us consider the  warped five-dimensional  Friedmann-Lema\^{\i}tre-Robertson-Walker (FLRW) model in cylindrical polar coordinates
\begin{eqnarray}
ds^2 = {\cal W}^2\Bigl[e^{2(\gamma-\psi)}(-dt^2+ dr^2)+e^{2\psi}dz^2+r^2 e^{-2\psi}d\varphi^2\Bigr]+ dy^2. \label{eqn1}
\end{eqnarray}
The function ${\cal W}$ is the warp factor and $y$ the extra (bulk) dimension. Here $\psi$ and $\gamma$ are  functions of $(t,r)$, while ${\cal W}$ is a function of $(t,r,y)$. Our 4-dimensional brane is located at $y=0$. All standard model fields resides on the brane, while gravity can propagate into the bulk.
We consider a scalar-gauge field on the brane in the form\cite{garfinkle85}
\begin{eqnarray}
\Phi=\eta X(t, r)e^{i\varphi},\qquad A_\mu =\frac{1}{ \epsilon }\bigl[P(t, r)-1\Bigr]\nabla_\mu\varphi, \label{eqn2}
\end{eqnarray}
with $\eta$ the vacuum expectation value of the scalar field and $\epsilon$ the coupling constant. As potential we take the well-known "mexican hat" potential $V(\Phi)=\frac{1}{8}\beta(\Phi^2-\eta^2)^2 $.
From the Einstein equations on the 5-dimensional spacetime one obtains a solution for the warp factor ${\cal W}$\cite{slagter15}
\begin{eqnarray}
{\cal W}=\frac{e^{\sqrt{- \tfrac{1}{6} \Lambda_5}(y- y_0)}}{\alpha\sqrt{ r}} \sqrt{\Bigl(d_1 e^{\alpha t}-d_2e^{-\alpha t}\Bigr)\Bigl(d_3 e^{\alpha r}-d_4e^{-\alpha r}\Bigr)}, \label{eqn3}
\end{eqnarray}
with $d_i$ and $\alpha$ some constants and $\Lambda_5$ the bulk cosmological constant. The first term in Eq.(\ref{eqn3}) is just the warp factor of the Randall-Sundrum model. The second term modifies the effective
4D Einstein equations.
The Einstein field equations induced on the brane can be derived using the Gauss-Codazzi equations and the Israel-Darmois junction conditions. The modified Einstein equations become\cite{sasaki00}
\begin{eqnarray}
{^{4}\!G}_{\mu\nu}=-\Lambda_{eff}{^{4}\!g}_{\mu\nu}+\kappa_4^2 {^{4}\!T}_{\mu\nu}+\kappa_5^4{\cal S}_{\mu\nu}-{\cal E}_{\mu\nu}, \label{eqn4}
\end{eqnarray}
with ${^{4}\!G}_{\mu\nu}$ the Einstein tensor calculated on the brane metric ${^{4}\!g}_{\mu\nu}= {^{5}\!g}_{\mu\nu}-n_\mu n_\nu$ and $n_\mu$ the unit vector normal to the brane.
In Eq.(\ref{eqn4}) the effective cosmological constant $\Lambda_{eff}=\frac{1}{2}(\Lambda_5+\kappa_4^2\Lambda_4)=\frac{1}{2}(\Lambda_5+\frac{1}{6}\kappa_5^4\Lambda_4^2)$ and $\Lambda_4$ is the vacuum energy in the brane (brane tension).  The latter equality sign is a consequence of the relation between the 4- and 5-dimensional Planck mass in the braneworld approach, $\kappa_5^4=6\frac{\kappa_4^2}{\Lambda_4}$. If in addition the brane tension is related to the 5-dimensional coupling constant and the cosmological constant by $ \frac{1}{6}\Lambda_4^2\kappa_5^4=-\Lambda_5$, then $\Lambda_{eff}=0$ and we are dealing with the RS-fine tuning condition\cite{randall99}.
The first correction term ${\cal S}_{\mu\nu}$ in Eq.(\ref{eqn4}) is  the quadratic term in the energy-momentum tensor arising
from the extrinsic curvature terms in the projected Einstein tensor
\begin{flalign}
{\cal S}_{\mu\nu}=\tfrac{1}{12}{^{4}\!T}{^{4}\!T}_{\mu\nu}-\tfrac{1}{4}{^{4}\!T}_{\mu\alpha}{^{4}\!T}^\alpha_\nu
+\tfrac{1}{24}{^{4}\!g}_{\mu\nu}\Bigl[3{^{4}\!T}_{\alpha\beta}{^{4}\!T}^{\alpha\beta}-{^{4}\!T}^2\Bigr]. \label{eqn5}
\end{flalign}
The second correction term ${\cal E_{\mu\nu}}$ in Eq.(\ref{eqn4})is given by
\begin{equation}
{\cal E}_{\mu\nu}={^{5}\!C}_{\alpha\gamma\beta\delta}n^\gamma n^\delta {^{4}\!g}_\mu^\alpha {^{4}\!g}_\nu^\beta, \label{eqn6}
\end{equation}
and is a part of the 5D Weyl tensor and carries information of the gravitational field outside the brane and is constrained by the motion of the matter on the brane, i.e., the Codazzi equation.
The scalar-gauge field equation becomes\cite{garfinkle85}
\begin{flalign}
D^\mu D_\mu\Phi =2\frac{dV}{d\Phi^*}, \qquad{^{4}\!\nabla}^\mu F_{\nu\mu}=\tfrac{1}{2}i\epsilon\Bigl(\Phi(D_\nu\Phi)^*-\Phi^* D_\nu \Phi \Bigr), \label{eqn7}
\end{flalign}
with $D_\mu \Phi \equiv {^{4}\!\nabla}_\mu \Phi +i\epsilon A_\mu\Phi, {^{4}\!\nabla}_\mu$ the covariant derivative with respect to ${^{4}\!g}_{\mu\nu}$, $\epsilon$ the gauge coupling constant  and the star represents the complex conjugated. $F_{\mu\nu}$
is the Maxwell tensor.

From Eq.(\ref{eqn4}) together with the matter field equations Eq.(\ref{eqn7}), one obtains a set of partial differential equations, which can be solved numerically\cite{slagter15}. Because gravity can propagate in the bulk, the cosmic string can build up a huge mass per unit length ( or angle deficit) $G\mu \sim1$ by the warp factor and can induce massive KK-modes felt on the brane, while the manifestation in the brane will be warped down to GUT scale, consistent with observations. Disturbances in the spatial components of the stress-energy tensor cause cylindrical symmetric waves, amplified due to the presence of the bulk space and warp factor. They could survive the natural damping due to the expansion of the universe. These disturbances could have a profound influence on the expansion of the universe. There could even be a "self-acceleration" without the need of an effective brane cosmological constant\cite{roy07}.

Besides the numerical solutions of the field equations, one should like to find an approximate wave solution where one can recognize the nonlinear features.  In order to keep track of  of the different orders of approximation, we will apply a multiple-scale analysis in the next section.

\section{Nonlinear wave approximation}\label{MS}
A linear approximation of  wave-like solutions of the Einstein equations is not adequate in the case of high energy or strong curvature.
There is a powerful approximation method to study non-linear gravitational waves without any averaging scheme. The method is  called a "two-timing" or "multiple-scale" method, because one considers the relevant fields $V_i$ in point ${\bf x}$ on a manifold M
dependent  on different scales $({\bf x}, \xi, \chi , ...)$\cite{choquet69,choquet77,slagter86}:
\begin{equation}
V_i=\sum_{n=0}^{\infty} \frac{1}{\omega^n} F_i^{(n)}({\bf x},\xi,\chi ,...).\label{eqn8}
\end{equation}
Here $\omega$ represents a dimensionless parameter, which will be large (the "frequency", $\omega>>1$). So $\frac{1}{\omega}$ is a small expansion parameter.
Further, $\xi =\omega \Theta({\bf x})$, $\chi =\omega \Pi({\bf x}), ...$ and $\Theta, \Pi , ...$  scalar (phase) functions on M. The small parameter $\frac{1}{\omega}$ can be the ratio of the characteristic wavelength of the perturbation to the characteristic dimension of the background. On warped spacetimes it could also be the ratio of the extra dimension y to the background dimension or even both.
One is interested in an approximate solution of the metric and  matter fields. If one substitute the series  Eq.(\ref{eqn8}) into the field equations, one obtains a formal series where now n runs from $-m$ to $\infty$, with m a constant.
One says that Eq.(\ref{eqn8}) is an approximate wavelike solution of order n of the field equations if $ F_i^{(n)}({\bf x},\xi,\chi ,...)=0$ for all n.
The method is very useful when one encounters non-uniformity in a regular perturbation expansion, i.e., the appearance of secular terms. In general relativity, this will occur when high-frequency gravitational waves interact with the background metric or the curvature is strong due to the presence of compact objects.
On our 5D spacetime, we expand
{\small
\begin{eqnarray}
g_{\mu\nu}=\bar g_{\mu\nu}({\bf x})+ \frac{1}{\omega}h_{\mu\nu}({\bf x},\xi,\chi ,..)+\frac{1}{\omega^2}k_{\mu\nu}({\bf x},\xi,\chi ,..) + ..., \cr
A_\mu=\bar A_\mu ({\bf x})+\frac{1}{\omega}B_\mu ({\bf x},\xi ,\chi ,..) +\frac{1}{\omega^2}C_\mu ({\bf x},\xi,\chi ,..) +... ,\cr
\Phi=\bar\Phi({\bf x}) +\frac{1}{\omega}\Psi({\bf x}, \xi, \chi ,..)+\frac{1}{\omega^2}\Xi({\bf x}, \xi, \chi ,..)+...,\label{eqn9}
\end{eqnarray}}
with $\bar g_{\mu\nu}$ the background metric and $\bar\Phi, \bar A_\mu$ the background scalar and gauge fields.
Let us consider, for the time being, only rapid variation in the direction of $l_\mu$  transversal to the sub-manifold $\Theta$ = constant (One could also consider independent rapid variation transversal to the sub-manifold $\Pi$ = constant).
We can  now define
\begin{eqnarray}
\frac{d g_{\mu\nu}}{d x^\sigma}=g_{\mu\nu,\sigma}+\omega l_\sigma \dot g_{\mu\nu}\qquad g_{\mu\nu,\sigma}\equiv \frac{\partial g_{\mu\nu}}{\partial x^\sigma}\qquad
\dot g_{\mu\nu}\equiv \frac{\partial g_{\mu\nu}}{\partial \xi},\label{eqn10}
\end{eqnarray}
with $l_\mu \equiv \frac{\partial \Theta}{\partial x^\mu}$.
We expand the several relevant tensors, for example,
\begin{eqnarray}
\Gamma_{\mu\nu}^\alpha =\bar \Gamma_{\mu\nu}^\alpha +\Gamma_{\mu\nu}^{\alpha (0)} +\frac{1}{\omega}\Gamma_{\mu\nu}^{\alpha (1)}+... ,\label{eqn11}
\end{eqnarray}
\begin{eqnarray}
R^\sigma_{\mu\tau\nu}=\omega R^{(-1)\sigma}_{\mu\tau\nu}+\bar R^\sigma_{\mu\tau\nu}+R^{(0)\sigma}_{\mu\tau\nu} +.... ,\label{eqn12}
\end{eqnarray}
with
\begin{equation}
\Gamma^{\sigma (0)}_{\mu\nu}=\tfrac{1}{2}\bar g^{\beta\sigma}\bigl( l_\mu \dot h_{\beta\nu}+l_\nu\dot h_{\beta\mu}-l_\beta \dot h_{\mu\nu}\bigr),\label{eqn13}
\end{equation}
\begin{flalign}
\Gamma^{\sigma (1)}_{\mu\nu}=\tfrac{1}{2}\Bigl(h^\sigma_{\mu :\nu}+h^\sigma_{\nu :\mu}-h_{\mu\nu}^{: \sigma}\Bigr)
-\tfrac{1}{2}\Bigl(l_\nu \dot k_\mu^\sigma + l_\mu \dot k_\nu^\sigma -l^\sigma \dot k_{\mu\nu}\Bigr)-h_\rho^\sigma\Gamma^{\rho (0)}_{\mu\nu},\label{eqn14}
\end{flalign}
where the colon represents the covariant derivative with respect to the ${^{4}\!\bar g_{\mu\nu}}$.
These expressions can also be calculated on ${^{5}\!g}_{\mu\nu}$. We substitute  the expansions into the effective brane Einstein equations Eq.(\ref{eqn4}) and subsequently put equal zero the various powers of $\omega$. We then obtain a system of partial differential equations for the fields $\bar g_{\mu\nu}, h_{\mu\nu}, k_{\mu\nu}$ and the scalar gauge fields $\bar \Phi, \Psi, \Xi, \bar A_\mu, B_\mu \equiv [B_0,B_1,0,B,0]$ and $C_\mu$. The perturbations can be $\varphi$-dependent.
The $\omega^{(-1)}$ Einstein equation becomes
\begin{equation}
{^{4}\!G_{\mu\nu}^{(-1)}}=-{\cal E}_{\mu\nu}^{(-1)}\label{eqn15},
\end{equation}
and the $\omega^{(0)}$ equation
\begin{flalign}
{^{4}\!\bar G_{\mu\nu}}+{^{4}\!G_{\mu\nu}^{(0)}}=-\Lambda_{eff}{^{4}\!\bar g_{\mu\nu}}
+\kappa_4^2 \bigl({^{4}\!\bar T_{\mu\nu}}+{^{4}\!T_{\mu\nu}^{(0)}}\bigr)
+\kappa_5^4\bigl(\bar {\cal S}_{\mu\nu}+{\cal S}_{\mu\nu}^{(0)}\bigr)-\bigl(\bar{\cal E}_{\mu\nu}+{\cal E}_{\mu\nu}^{(0)}\bigr),\label{eqn16}
\end{flalign}
The contribution from the bulk space, ${\cal E}_{\mu\nu}^{(-1)}$, must be calculated with the 5D Riemann tensor
\begin{equation}
{^{5}\!R^{(-1)\sigma}_{\mu\tau\nu}}=l_\tau {^{5}\!\dot \Gamma^{(0)\sigma}_{\mu\nu}-l_\nu }{^{5}\!\dot\Gamma^{(0)\sigma}_{\mu\tau}}.\label{eqn17}
\end{equation}
If we consider $l_\mu l^\mu =0$, i.e., the eikonal equation, then one obtains from Eq. (\ref{eqn15})
\begin{equation}
l^\alpha\bigl(\ddot h_{\alpha\nu}-\tfrac{1}{2}\bar g_{\alpha\nu}\ddot h\bigr)=0,\label{eqn18}
\end{equation}
which in other contexts is used as gauge conditions. It turns out that the contribution from the ${\cal E}_{\mu\nu}^{(-1)}$ don't change this conditions  on $h_{\mu\nu}$, if we take
$h_{12}=\frac{1}{2}(h_{11}+h_{22})$, which is a pleasant fact.
Let us consider as a simplified case $l_\mu =[1,1,0,0,0]$. Then we obtain from the gauge condition Eq.(\ref{eqn18}) that only $h_{11}, h_{22}, h_{44}, h_{55}, h_{13}, h_{14}, h_{15}, h_{34}, h_{45}$ and $h_{35} $ survive.
If $l_\mu l^\mu \neq 0$, one proves in the 4D case that the $h_{\mu\nu}$ arises from a coordinate transformation and $\Theta=cst$ is not a wavefront of the background.
Let us consider the zero-order  Eq. (\ref{eqn16}). The most important contribution comes from
\begin{flalign}
{\cal E}_{\mu\nu}^{(0)}=n^\gamma n^\delta {^{4}\!g_\mu^\alpha}{^{4}\!g_\nu^\beta}\Bigl[{^{5}\!R^{(0)}_{\alpha\gamma\beta\delta}}
-\tfrac{1}{3}\Bigl({^{5}\!\bar g_{\alpha\gamma}} {^{5}\!R^{(0)}_{\delta\beta}}-{^{5}\!\bar g_{\alpha\delta}} {^{5}\!R^{(0)}_{\gamma\beta}}
-{^{5}\!\bar g_{\beta\delta}} {^{5}\!R^{(0)}_{\gamma\alpha}}
 +{^{5}\!\bar g_{\beta\delta}} {^{5}\!R^{(0)}_{\gamma\alpha}}\Bigr)
+\tfrac{1}{12}\Bigl({^{5}\!\bar g_{\alpha\gamma}}{^{5}\!\bar g_{\delta\beta}}-{^{5}\!\bar g_{\alpha\delta}}{^{5}\!\bar g_{\gamma\beta}}\Bigr){^{5}\!R}\Bigr].\label{eqn19}
\end{flalign}
One also needs  the Ricci tensor ${^{4}\!R^{(0)}_{\mu\nu}}$ in  Eq.(\ref{eqn16}), which is given by ( for $\sigma =\tau$)
\begin{flalign}
{^{4}\!R^{\sigma (0)}_{\mu\tau\nu}}={^{4}\!\Gamma^{\sigma (0)}_{\mu\nu :\tau}}-{^{4}\!\Gamma^{\sigma (0)}_{\mu\tau :\nu}}+ {^{4}\!\Gamma^{\rho (0)}_{\mu\nu}}{^{4}\!\Gamma^{\sigma (0)}_{\rho\tau}}
-{^{4}\!\Gamma^{\rho (0)}_{\mu\tau}}{^{4}\!\Gamma^{\sigma (0)}_{\rho\nu}}+l_\tau {^{4}\!\dot \Gamma^{\sigma (1)}_{\mu\nu}}-l_\nu {^{4}\!\dot \Gamma^{\sigma (1)}_{\mu\tau}}.\label{eqn20}
\end{flalign}
From the Einstein equations Eq.(\ref{eqn16}), one can deduce  a set of partial differential equations (PDE's) when one imposes additional gauge conditions. As a simplified model, we take $h_{22}=-h_{11}, h_{34}=h_{35}=h_{45}=h_{14}=h_{15}=0$ (leaving 4
independent $h_{\mu\nu}$ terms), we have 7 unknown functions for the background and first order perturbations: $\bar W_1, \bar\psi, \bar \gamma, \dot h_{13}, \dot h_{11}, \dot h_{44}$ and  $\dot h_{55}$.
One  can also  integrate the equation Eq.(\ref{eqn16}) with respect to $\xi$. If we suppose that the perturbations are periodic in $\xi$, we then obtain the Einstein equations with back-reaction terms:
\begin{flalign}
{^{4}\!\bar G_{\mu\nu}}=\kappa_4^2{^{4}\!\bar T_{\mu\nu}}+\kappa_5^4 \bar {\cal S}_{\mu\nu}-\bar{\cal E}_{\mu\nu}
+\frac{1}{\tau}\int \Bigl(\kappa_4^2T_{\mu\nu}^{(0)}+\kappa_5^4S_{\mu\nu}^{(0)}-{^{4}\!G_{\mu\nu}^{(0)}}-{\cal E}_{\mu\nu}^{(0)}\Bigr ) d\xi ,\label{eqn21}
\end{flalign}
where we took $\Lambda_{eff}=0$ for the RS fine-tuning and  $\tau$ de period of the high-frequency components.
One can say that the term $-\int {\cal E}_{\mu\nu}^{(0)}d\xi$ in Eq.(\ref{eqn21})is the KK-mode contribution of the perturbative 5D graviton. It is an extra back-reaction term, which contain $\dot h_{55}$ amplified by the warp factor and with opposite sign with repect to the $\kappa_4^2$-term. So it can play the role of an effective cosmological constant.
By substituting back these equations into the original equations, one  gets propagations equations for the first order perturbations. In this way we obtain the set PDE's\cite{slagter15b}
\begin{flalign}
\partial^2_{tt}\bar W_1=-\partial^2_{rr}\bar W_1+\frac{2}{\bar W_1}(\partial_t\bar W_1^2+\partial_r\bar W_1^2)-
\bar W_1(\partial_t\bar \psi^2+\partial_r\bar \psi^2)
+\frac{\bar W_1}{r}(\partial_r\bar \gamma-\partial_t\bar\gamma)\cr
+2(\partial_r\bar W_1-\partial_t\bar W_1)(\partial_t\bar\psi-\partial_r\bar\psi
+\partial_r\bar\gamma-\partial_t\bar\gamma)
+2\bar W_1\partial_r\bar\psi\partial_t\bar\psi-4\frac{\partial_r\bar W_1\partial_t\bar W_1}{\bar W_1}
+ 2\partial_{rt}\bar W_1\cr
-\tfrac{3}{4}\kappa_4^2\Bigl(e^{2\bar\psi}\frac{(\partial_t\bar P-\partial_r\bar P)^2}{\bar W_1 r^2 \epsilon^2}
+\bar W_1(\partial_t\bar X-\partial_r\bar X)^2\Bigr),\label{eqn22}
\end{flalign}
 \begin{flalign}
\partial_{tt}\bar\psi=\partial_{rr}\bar\psi +\frac{\partial_r\bar\psi}{r}+\frac{2}{\bar W_1}(\partial_r \bar W_1\partial_r\bar\psi -\partial_t\bar W_1\partial_t\bar\psi)
-\frac{\partial_r\bar W_1}{r\bar W_1}
+\frac{3 e^{2\bar\psi}}{4\bar W_1^2r^2\epsilon^2}\kappa_4^2\Bigr(\partial_t\bar P^2-\partial_r\bar P^2\cr
-\bar W_1^2 \epsilon^2\bar X^2\bar P^2 e^{2\bar\gamma-2\bar\psi}\Bigr),\label{eqn23}
\end{flalign}
\begin{flalign}
\partial_{t}\bar\gamma =\partial_r\bar\gamma+\frac{1}{\partial_t\bar W_1-\partial_r\bar W_1-\frac{\bar W_1}{2r}}\Bigr\{\tfrac{1}{2}\bar W_1(\partial_t\bar\psi -\partial_r\bar\psi)^2
+\frac{\partial_r\bar W_1}{r}
-\partial_{tr}\bar W_1+\partial_{rr}\bar W_1 +\frac{2\partial_r\bar W_1\partial_t\bar W_1}{\bar W_1}\cr
+(\partial_r\bar W_1-\partial_t\bar W_1)(\partial_r\bar \psi-\partial_t\bar \psi)
-\frac{\partial_r\bar W_1^2+3\partial_t\bar W_1^2}{2\bar W_1}
+\kappa_4^2\frac{\bar W_1}{16}\Bigl(7\partial_t\bar X^2+5\partial_r\bar X^2-12\partial_r\bar X\partial_t\bar X\cr
+5e^{2\bar\gamma}\frac{\bar X^2\bar P^2}{r^2}+6e^{2\bar\psi}\frac{(\partial_r\bar P -\partial_t\bar P)^2}{\bar W_1^2r^2\epsilon^2}
+\bar W_1^2\beta e^{2\bar\gamma-2\bar\psi}(\bar X^2-\eta^2)^2\Bigr)\Bigr\},\label{eqn24}
\end{flalign}
\begin{flalign}
\partial_t\dot h_{13}=\partial_r\dot h_{13}+\ddot k_{13}-\ddot k_{23}+2\Bigl(\frac{\partial_t\bar W_1 -\partial_r\bar W_1}{\bar W_1}
+\partial_t\bar\psi -\partial_r\bar\psi\Bigr)\dot h_{13},\label{eqn25}
\end{flalign}
\begin{flalign}
\partial_t\dot h_{11}=\partial_r\dot h{11}+\frac{e^{2\bar\gamma}}{r^2}\Bigl(\partial_r\bar\psi-\partial_t\bar\psi-\tfrac{1}{2r}\Bigr)\dot h_{44}
+\tfrac{1}{2}(\ddot k_{22}+\ddot k_{11})-\ddot k_{12}+
\frac{2}{\bar W_1}\Bigl(\partial_t\bar W_1-\partial_r\bar W_1\cr
+\bar W_1(\partial_r\bar\psi-\partial_t\bar\psi+\partial_t\bar\gamma-\partial_r\bar\gamma)\Bigr)\dot h_{11}\cr
+\tfrac{1}{2}e^{2\bar\gamma-2\bar\psi}\bar W_1^2\Bigr(\tfrac{1}{2r}+\frac{\partial_r\bar W_1-\partial_t\bar W_1}{\bar W_1}\Bigr)\dot h_{55}
+\kappa_4^2e^{2\bar\gamma-2\bar\psi}\bar W_1^2(\partial_t\bar X-\partial_r\bar X) \dot\Psi\cos\varphi,\label{eqn26}
\end{flalign}
\begin{flalign}
\partial_\varphi\Bigl(\dot h_{11}+\frac{e^{2\bar\gamma}}{r^2}\dot h_{44}-\bar W_1^2 e^{2\bar\gamma-2\bar\psi}\dot h_{55}\Bigr)+\ddot k_{24}-\ddot k_{14}=
-2\kappa_4^2\bar X\bar Pe^{2\bar\gamma-2\bar\psi}\bar W_1^2\sin\varphi\dot\Psi,\label{eqn27}
\end{flalign}
\begin{equation}
\partial_t\dot h_{55}-\partial_r\dot h_{55}=0,\label{eqn28}
\end{equation}
\begin{flalign}
\partial_t\dot h_{44}=\partial_r\dot h_{44}+\Bigl(2\partial_r\bar\psi-2\partial_t\bar\psi-\tfrac{3}{2r}
+\frac{\partial_t\bar W_1-\partial_r\bar W_1}{\bar W_1}\Bigr)\dot h_{44}
+\frac{\kappa_4^2}{\epsilon}(\partial_r\bar P-\partial_t\bar P)\dot B\cr
+\tfrac{1}{2}\bar W_1^2 r^2e^{-2\bar\psi}\Bigl(\partial_t\bar\psi-\partial_r\bar\psi+\tfrac{1}{2r}\Bigr)\dot h_{55}.\label{eqn29}
\end{flalign}
We notice that in our simplified case of radiative coordinates $\Theta(x_\mu)=t+r$, the equations for the background metric separates from the perturbations. So this example is very suitable to investigate the perturbation equations.
For the first order gauge field perturbation $B_\mu$ we used the condition $l^\mu B_\mu =0$, which is a consequence, as we will see,  of the gauge field equations. So $B_\mu$ can be parameterized as $B_\mu =[B_0,B_0,0,B,0]$.
The propagation equation for $\dot h_{55}$ yields $\dot h_{55}=F_1(t+r)F_2(\varphi,y,\xi)$, which is expected, because the brane part of $\dot h_{55}$ must be separable from the bulk part.
We omitted for the time being, the $\kappa_5^4$ contribution.

It is manifest that to zero order there is an interaction between the high-frequency perturbations from the bulk, the matter fields on the brane and the evolution of $\dot h_{ij}$, also found in the numerical solution {\cite{slagter15}. We observe again that the bulk contribution $\dot h_{55}$ is amplified by $\bar W_1^2$. It is a reflection of the massive KK modes felt on the brane.
The contribution of $\dot h_{55}$ in Eq.(29) disappears when $\partial_t\bar\psi-\partial_r\bar\psi+\frac{1}{2r}=0$. In the static case this results in a solution $\bar\psi= a\log r +b$ (a=1/2  in our case). This solution is of less physical significance because a distant test particle in this field will be repelled from the cylinder for $0<a<1$\cite{stachel68}.
The equations for the matter fields can be obtained in a similar way. The equation for the background $\bar \Phi$ becomes
\begin{flalign}
\bar D^\alpha\bar D_\alpha\bar\Phi-\frac{1}{2}\beta\bar\Phi(\bar\Phi\bar\Phi^*-\eta^2)
=\frac{1}{\tau}\int\Bigl(h^{\mu\nu}l_\mu l_\nu\ddot\Psi +\bar g^{\mu\nu}\Gamma^{\alpha(0)}_{\mu\nu}\dot\Psi\Bigr)d\xi.\label{eqn30}
\end{flalign}
The equation for $\bar A_\mu$ is the same as in the unperturbed situation. For the first order perturbations we obtain ( for $l^\alpha C_\alpha =0$)
\begin{equation}
\partial_t \dot\Psi=\partial_r\dot\Psi+\frac{\dot\Psi}{\bar W_1}(\partial_r\bar W_1 -\partial_t\bar W_1)+\frac{1}{2r}\dot\Psi,\label{eqn31}
\end{equation}
\begin{flalign}
\partial_t\dot B=\partial_r\dot B+\Bigl(\partial_r\bar\psi-\partial_t\bar\psi-\frac{1}{2r}\Bigr)\dot B
+\frac{e^{2\bar\psi}(\partial_t\bar P-\partial_r\bar P)}{2r^2\bar W_1^2 \epsilon}\dot h_{44},\label{eqn32}
\end{flalign}
\begin{equation}
\partial_t \dot B_0 =\partial_r \dot B_0-\frac{e^{2\bar\gamma}}{r^2}\partial_\varphi \dot B -\epsilon e^{2\bar\gamma-2\bar\psi}\dot\Psi \bar X\bar W_1^2\sin\varphi.\label{eqn33}
\end{equation}
For these matter field equations one needs the condition $l^\alpha \bar A_\alpha =0$, otherwise the real and imaginary parts of $\dot \Psi$ interact as the propagation progresses.
From Eq.(\ref{eqn26}), Eq.(\ref{eqn27}) and Eq.(\ref{eqn33}) we observe  on the right hand side  $\varphi$-dependent terms, amplified by $W_1^2$. So the approximate wave solution is no longer axially symmetric, also found by \cite{choquet69}. After integration with respect to $\varphi$, we obtain from Eq.(\ref{eqn27}) ( for $k_{14}=k_{24}$)
\begin{flalign}
\dot h_{11}=e^{2\bar\gamma -2\bar\psi}\bar W_1^2\dot h_{55}-\frac{e^{2\bar\gamma}}{r^2}\dot h_{44}
-2\kappa_4^2e^{2\bar\gamma-2\bar\psi}\bar X\bar P\bar W_1^2\int(\dot\Psi\sin\varphi) d\varphi.\label{eqn34}
\end{flalign}
This means that the $(r,r)$ first order disturbance $\dot h_{22}$ ($\dot h_{22}=-\dot h_{11}$)) could have its maximum for fixed angle $\varphi$ amplified by the warp factor $W_1^2$.
If we choose for example
$\Psi=\cos\varphi \tilde\Psi(t,r,\xi)$, then the last term in Eq.(\ref{eqn34}) becomes $\kappa_4^2\bar X\bar P e^{2\bar\gamma-2\bar\psi}\bar W_1^2\cos 2\varphi \dot{\tilde\Psi}$, which has  two extremal values  on $[0,\pi]$ $\mod(\frac{1}{2}\pi )$. The energy-momentum tensor component ${^{4}\!T}_{rr}^{(0)}$ is
\begin{flalign}
{^{4}\!T}_{rr}^{(0)}=\dot\Psi^2+\dot\Psi(\partial_t\bar X+\partial_r \bar X)\cos  \varphi 
+\frac{e^{2\bar\psi}}{\bar W_1^2 r^2 \epsilon}\Bigr(\epsilon \dot B^2+\dot B(\partial_r\bar P+\partial_t\bar P)\Bigr)\label{eqn35}
\end{flalign}
This angle-dependency could be an explanation of the recently found spooky alignment of the rotation axes of quasars over large distances in two perpendicular directions.

The next step is to investigate the higher order equations in $\omega$ , which will provide the propagation equations of $k_{\mu\nu}$ and back-reaction terms in the background field equations Eq.(\ref{eqn22}), (\ref{eqn23})
 and Eq.(\ref{eqn24}). In this way, one can construct an approximate solution of the Einstein and scalar-gauge field equations and one can keep track of the different orders of perturbations.
\section{Conclusions}
A nonlinear approximation of the field equations of the coupled Einstein-scalar-gauge field equations on a warped 5D spacetime is investigated.  To zeroth order in the expansion parameter it is found that the evolution
of the perturbations on the brane is triggered by the electric part of the 5D Weyl tensor and carries information of the gravitational field outside the brane. The warpfactor in the nominator in front of the bulk contributions will cause a huge disturbance on the brane and could act as dark energy. It turns out that the first order disturbances are no longer axially symmetric. This means that wave-like disturbances in the energy-momentum tensor components can have preferred $\varphi$ directions  perpendicular to each other. This could be an explanation of the alignment of the preferred directions of the quasar polarization axes.

 {\color{myaqua}

}}

\end{document}